\documentclass{nicso}
\usepackage[latin1]{inputenc}
\usepackage{color}
\usepackage{graphicx}
\usepackage{longtable}
\usepackage{multirow}
\usepackage{amsmath}
\usepackage{url}

\definecolor{color1}{rgb}{0.000,0.000,0.000}
\definecolor{color2}{rgb}{0.000,0.000,1.000}
\definecolor{color3}{rgb}{0.000,1.000,1.000}
\definecolor{color4}{rgb}{0.000,1.000,0.000}
\definecolor{color5}{rgb}{1.000,0.000,1.000}
\definecolor{color6}{rgb}{1.000,0.000,0.000}
\definecolor{color7}{rgb}{1.000,1.000,0.000}
\definecolor{color8}{rgb}{1.000,1.000,1.000}
\definecolor{color9}{rgb}{0.000,0.000,0.502}
\definecolor{color10}{rgb}{0.000,0.502,0.502}
\definecolor{color11}{rgb}{0.000,0.502,0.000}
\definecolor{color12}{rgb}{0.502,0.000,0.502}
\definecolor{color13}{rgb}{0.502,0.000,0.000}
\definecolor{color14}{rgb}{0.502,0.502,0.000}
\definecolor{color15}{rgb}{0.502,0.502,0.502}
\definecolor{color16}{rgb}{0.753,0.753,0.753}
\definecolor{color17}{rgb}{1.000,1.000,1.000}

\begin{document}

\title*{CHAC. A MOACO Algorithm for Computation of  Bi-Criteria Military Unit 
Path in the Battlefield}

\titlerunning{CHAC. A MOACO Algorithm...}

\author{A.M. Mora\inst{1} \and J.J. Merelo\inst{1} \and
C. Millan\inst{2} \and  J. Torrecillas\inst{2}  \and   
J.L.J. Laredo\inst{1}}
\authorrunning{A. M. Mora et al.}

\institute{
Departamento de Arquitectura y Tecnología de Computadores\\
University of Granada (Spain) \\
\email{\{amorag,jmerelo,juanlu\}@geneura.ugr.es}
\and 
Mando de Adiestramiento y Doctrina\\
Spanish Army\\
\email{\{cmillanm, jtorrelo\}@et.mde.es}
}

\maketitle

\begin{abstract}
In this paper we propose a Multi-Objective Ant Colony Optimization
(MOACO) algorithm called CHAC, which has been designed to solve the
problem of finding the path on a map (corresponding to a simulated
battlefield) that minimizes resources while maximizing safety. CHAC
has been tested with two different 
state transition rules: an aggregative function that combines the heuristic and pheromone
information of both objectives and a second one that is based on the
dominance concept of multiobjective optimization problems. These rules
have been evaluated in several different situations (maps with
different degree of difficulty), and we have found
that they yield better results than a greedy algorithm (taken as
baseline) in addition to a {\em military} behaviour that is also
better in the tactical sense. The aggregative function, in general,
yields better results than the one based on dominance. 
\end{abstract}

\section{Introduction}

Prior to any combat manoeuvre, the unit commander must plan the best
path to get to an position that is advantageous over the enemy in the
tactical sense. This decision is conditioned by two criteria, speed
and safety, which he must evaluate. The choice of a safe path is made
when the enemy forces situation is not known, so the unit must move
through hidden zones in order to avoid detection, which may correspond
to a very long, and thus slower, path. On the other hand, the choice
of a fast path is made when the unit is going to attack the enemy or
if there are few hidden (safe) zones in the terrain and going through
it may produce a lot of casualties. In any case, the computation of
the best itinerary for the unit to reach the objective point is a very
important tactical decision. We will try to make this decision in the
paper by solving the what we have called the military unit path
finding problem.

The military unit path finding problem is similar to the common 
path finding problem with some additional features and restrictions. 
It intends to find the best path from an origin to a destination 
point in the battlefield but keeping a balance between route speed 
and safety.  The unit has an energy (health) and a resource level which 
are consumed when the unit moves through the path depending on 
the kind of terrain and difference of height, so the problem objectives 
are adapted to minimize the consumption of resources (which usually 
means walking as short/fast as possible) and the consumption 
of energy. In addition, there might be some enemies in the map 
(battlefield) which could shoot their weapons against the unit.

To solve the problem an Ant Colony System (ACS) algorithm has 
been adapted. It is a type of ACO \cite{dori99a,dori2003}, which
offers more control over the exploration and exploitation parts of
search. In addition, some  
features to deal with the two objectives of the problem have 
been added (in \cite{Coello2002} it can be found a survey of MO algorithms), 
so it is a MOACO algorithm (paper \cite{bicriteria} presents a review of some 
of them). This problem had been solved so far using classical techniques 
like branch and bound or dynamic programming, which do not usually 
scale well with size, but to the best of our knowledge find no 
one that treat the problem as a multiobjective and solve it with 
a MOACO.

The rest of the paper is organized as follows: next section is devoted
to a presentation of the problem we want to solve, followed in section
\ref{sec:chac} by a presentation of the CHAC algorithm. Experiments
and results are presented in section \ref{sec:exp}, and conclusions
are drawn in section \ref{sec:conc}, along with the future lines of
development of this work.

\section{The Problem}

Our objective in this paper is to give the unit commander, or to a 
simulated unit in a combat training simulator, a tactical decision 
capability so that it will be able to calculate the best path 
to its target point in the battlefield considering the same factors 
that a commander would. The battlefield has been modeled as a square
cell grid, every cell corresponding  to a 500x500 meter zone in the
real world.

The speed in a path is associated with the unit resources consumption 
because we have assigned a penalization to every cell related 
to the `difficulty' of going through it and we have called this 
penalization resource cost of the cell. So, going through cells 
with more resource cost is more difficult and so more slow. Due 
to this justification, we refers to fast paths or paths with 
small resource cost.

Thus, the problem unit has two properties, a number of \textit{energy} 
(which represent global health of unit soldiers or status of the unit 
vehicles) points and a number of \textit{resource} (which represent 
unit supplies such as fuel, food and even moral) points. Its objective 
is to get to a target point with the maximum level in both properties.

The unit energy level is decremented in every cell of the path 
(the company depletes its human resources or vehicles suffer 
damage), which have a penalization called \textit{no combat casualties}, 
depending on its type. In addition there is an extra penalization 
due to the impact of an enemy weapon in the cell. This value 
is calculated as combination of three other, the probability 
of enemy shoots, the probability of impact in the unit, and the 
damage it would produce to the unit.

Moreover every cell has an assigned resources penalty depending 
on the type of terrain. There is an extra penalty when the unit 
goes from a cell to other with different height (if it goes down, 
the penalty is small than if it goes up).

Besides \textit{cost in energy}, and \textit{cost in resources}, cells have 
the following properties:
\begin{itemize}
\item\textit{Type}: there are four kinds of cells: normal (flat terrain), 
forest, water and obstacle, three different types of terrain and obstacle 
which means a cell that the unit cannot going through. There are 
several \textit{subtypes}: it can be an enemy unit position, problem unit 
position, or the target point;  it can also be affected by enemy 
fire (with one hundred levels of impact) or be lethal for the 
problem unit.
\item \textit{Height}: an integer number (between -3 and 3) which represents 
a level of depth or height of a cell.
\end{itemize}

\begin{figure}[htbp]
\begin{center}
\includegraphics[scale=0.4]{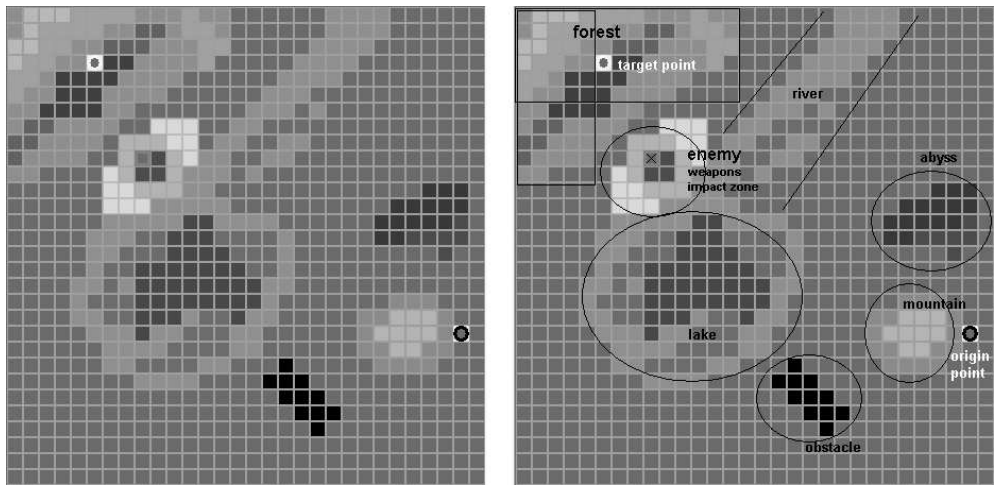}
\caption{Map Example. It is a 30x30 map which shows some of the types
  and subtypes above mentioned. In left, and explanation of cells
  types is showed, because it is difficult to differ between different
  kinds of cells due to the black and white image. The example
  includes the problem unit (circle with black border), an enemy unit
  (square with gray border and mark with 'X'), the zone affected by
  enemy weapons (in different shades of light gray depending on the
  associated damage) surrounding the enemy and the objective point for
  the unit (circle with light gray border).  
The different shades in the same color models height (light color) 
and depth (dark color).\label{fig:example}}
\end{center}
\end{figure}

We have implemented an application in Delphi 7 for Windows XP 
in order to create the problem scenarios (battlefields) and also 
to visualize the solutions obtained by CHAC. This application 
is available under request. Figure \ref{fig:example} shows an example of battlefield. 
It includes all typical features: enemy unit, obstacles, forest 
and water filled terrain accidents.

This implementation includes some constraints: problem and enemy 
units fill up exactly one cell (which corresponds with their 
size in real world). The problem unit as well as the target point must 
be located on the map; the presence of the enemy is optional. The 
unit can go only once through a cell and cannot go through cells 
occupied by an enemy or an obstacle, or cells with a cost in 
resources or energy bigger than what the unit has available. Some 
rules also apply to the \textit{line of sight}, so that it behaves 
as realistically as possible.

\section{The CHAC Algorithm}
\label{sec:chac}

CHAC means \textit{Compañía de Hormigas Acorazadas} or Armoured Ant 
Company; we have chosen this (admittedly a bit tongue-in-cheek) name
to relate the ACO algorithms with the military scope  
of this application. It is an ACS adapted to deal with several 
objectives, that is, a MOACO or Multi-objective Ant Colony
Optimization algorithm. 

In order to approach it via an  ACO algorithm, the problem must be
transformed into a graph with weighted edges. Every cell in the map is considered 
a node in the graph with 8 edges connect it to every one of its 
neighbours (except in border cells). To deal with two objectives, 
there are two weights in every edge related to the cost of energy 
and resources, that is, a cost related to the energy expenses (cost 
assigned to the destination node of the edge) and other related 
to the consumption of resources if the unit moves from one cell 
to its neighbour following that edge (which depends on the types 
of both nodes and the height difference between them).

The algorithm implemented within CHAC is constructive which means 
that in every iteration, every ant builds a complete solution, if 
possible (there are some constraints, for example the unit cannot go 
through one node twice in the same path), by travelling through 
the graph. In order to guide this movement, the algorithm uses information 
of two kinds: pheromone and heuristic, that will be combined.


Since it is a multiobjective problem, CHAC uses two pheromone 
matrices, one per objective, and two heuristics functions (also 
matrices), following the BicriterionAnt algorithm designed by 
Iredi et al. in \cite{IMM2001}. However, we decided to use an Ant Colony 
System (ACS) instead of an Ant System to have better control in 
the balance between exploration and exploitation by using the parameter \textit{q$_{0}$}. We have implemented two state transition rules (which means two different algorithms), first one similar to the Iredi's proposal 
and second one based on dominance of neighbours. The local and 
global updating formulas are based in the MACS-VRPTW algorithm 
proposed by Barán et al. in \cite{Baran}, with some changes due to the 
use of two pheromone matrices.

The objectives are named \textbf{\textit{f}}, minimization the resources 
consumed in the path (speed maximization) and \textbf{\textit{s}}, minimization 
the energy consumed in the path (safety).\\
\textit{The Heuristic Functions} try to guide search between 
the start and the target point considering the most important factors 
for every objective. So, for edge \textit{(i,j)} they are:

\begin{equation} \label{eq:nf}
\eta _{f} (i,j)=\frac{\omega _{fr} }{Cr(i,j)} +\frac{\omega _{fd}
}{Dist(j,T)} + \left(\omega _{fo} \cdot ZO(j)\right)
\end{equation}

\begin{equation} \label{eq:ns}
\eta _{s} (i,j)=\frac{\omega _{se} }{Ce(i,j)} +\frac{\omega _{sd}
}{Dist(j,T)} +\left(\omega _{so} \cdot ZO(j)\right)
\end{equation}

In Equation \ref{eq:nf}, \textit{Cr} is the resource cost when moving
from node \textit{i} to node \textit{j}, \textit{Dist} is the
Euclidean distance between two nodes (\textit{T} is the target node of
the problem) and \textit{ZO} is  
a score (between 0 and 1) to a cell being 1 when the cell is hidden to
all the enemies (or to all the cells in a radius when there are no
enemies) and decreasing exponentially when it is seen. \textit{$\omega
_{fr}$} , \textit{$\omega _{fd}$} and \textit{$\omega _{fo}$}  
are weights to assign relative importance to the terms in the 
formula. In this case, the most important term is the distance to 
target point because it searches for the fastest path. In second 
place is important the resources cost and only a little the hidden 
of the cell (it almost does not mind in a fast path).

In Equation \ref{eq:ns}, \textit{Ce} is the energy cost of moving to node \textit{j}, \textit{Dist} and \textit{ZO} is the same as the previous formula. \textit{$\omega _{se}$} , \textit{$\omega _{sd}$} and \textit{$\omega _{so}$}  
are again weights to assign relative importance to the terms in 
the formula, but in this case the most important are energy cost 
and hidden (both are to be considered in a safe path) and a little 
the distance to target point.

\textit{The Combined State Transition Rule} (CSTR) is a formula 
used to decided which node \textit{j} is the next in the construction 
of a solution (path) when the ant is at the node \textit{i}, it is the pseudo-random-proportional 
rule used in ACS, but adapted to deal with a two objectives problem 
by combining the heuristic and pheromone information of both 
of them:

If (q $\le$ q0)

\begin{equation} \label{eq:rtj}
j=\arg \max  _{j\in N_{i} } \, \left\{\tau _{f} (i,j)^{\alpha \cdot \lambda }
\cdot \tau _{s} (i,j)^{\alpha \cdot (1-\lambda )} \cdot \eta _{f}
(i,j)^{\beta \cdot \lambda } \cdot \eta _{s} (i,j)^{\beta \cdot (1-\lambda
)} \right\}
\end{equation}

Else

\begin{equation} \label{eq:rtp}
P(i,j)=\left\{ \begin{array}{ll} 
\frac { \displaystyle \tau _{f} (i,j)^{\alpha \cdot \lambda }\cdot \tau _{s} (i,j)^{\alpha \cdot (1-\lambda )} \cdot \eta _{f}(i,j)^{\beta \cdot \lambda } \cdot \eta _{s} (i,j)^{\beta \cdot (1-\lambda)} } { \displaystyle \sum\limits_{u\in N_{i}} \tau _{f} (i,u)^{\alpha \cdot \lambda }\cdot \tau _{s} (i,u)^{\alpha \cdot (1-\lambda )} \cdot \eta _{f}
(i,u)^{\beta \cdot \lambda } \cdot \eta _{s} (i,u)^{\beta \cdot (1-\lambda
)} } & \quad if \, j\in N_{i} \\
\\
\\
0 & \quad otherwise\\
\end{array} \right.
\end{equation}

Where \textit{q}$_{0}$ $\in$ [0,1] is the standard ACS parameter, \textit{q} 
is a random value in [0,1]. \textit{$\tau _{f}$} and \textit{$\tau _{s}$} 
are the pheromone matrices and \textit{$\eta _{f}$} and \textit{$\eta _{s}$} 
are the heuristic functions for the objectives (Equations \ref{eq:nf} and 
\ref{eq:ns}). All these matrices have a value for every edge \textit{(i,j)}. \textit{$\alpha$} and \textit{$\beta$} are the usual weighting parameters and \textit{N$_i$} is the current feasible neighbourhood for the node \textit{i}. \\
\textit{$\lambda$} $\in$ (0,1) is a user-defined parameter which sets the importance of the objectives in the search (this is an application created for a military user who decides which objective has more priority), so for instance, if the user decides to search for the fastest path, \textit{$\lambda$} will take a value close to 1 if he wants the safest path, it has to be close to 
0. This value is kept during all the algorithm and for all the 
ants, unlike other bi-criterion implementations in which the parameter takes value 0 for first ant and it was growing for every ant until it takes 
a value 1 for the last one.

When an ant is building a solution path and it is placed at one 
node \textit{i}, if \textit{q}  $\le$ \textit{q}$_{0}$ the best neighbour \textit{j} is selected as the next (Equation \ref{eq:rtj}). Otherwise, the algorithm decides which node is the next by using a roulette considering \textit{P(i,j)} as probability for every feasible neighbour \textit{j} (Equation \ref{eq:rtp}).

\textit{The Dominance State Transition Rule} (DSTR) is based 
on the dominance concept (see reference \cite{Coello2002}), which is defined 
as follows (\textit{a} dominates \textit{b}):

\begin{equation} \label{eq:dom}
a\succ b\;\;if: \quad
\forall i\in 1,2,...,k \quad | \quad C_{i} (a)\leq C_{i} (b)\quad \wedge \quad \exists j\in 1,2,...k|C_{j} (a)<C_{j} (b)
\end{equation}

Where \textit{a} and \textit{b} are two different vectors of \textit{k} values 
(one per objective) and \textit{C} is a cost function for every component 
in the vector. If it intends to minimize the cost and Equation \ref{eq:dom} is true, then \textit{b} is dominated by \textit{a}.

So, in our problem there are two cost functions to evaluate the 
dominance between nodes. Actually these functions consider edges 
because they have assigned pheromone and heuristic information 
and combine them:

\begin{equation} \label{eq:cf}
C_{f} (i,j)=\tau _{f} (i,j)^{\alpha \cdot \lambda } \cdot \eta _{f}
(i,j)^{\beta \cdot \lambda }
\end{equation}

\begin{equation} \label{eq:cs}
C_{s} (i,j)=\tau _{s} (i,j)^{\alpha \cdot (1-\lambda )} \cdot \eta _{s}
(i,j)^{\beta \cdot (1-\lambda )}
\end{equation}
\\
In addition there is a function which uses Equations \ref{eq:cf} and \ref{eq:cs}:

\begin{equation} \label{eq:fundom}
D(i,j,u)=\left\{ \begin{array}{ll} 
1 & \quad if \: (i,j)\succ(i,u)\\
\\
0 & \quad otherwise
\end{array} \right.
\end{equation}
\\
At last, the dominance state transition rule is as follows:

If (q $\le$ q0)

\begin{equation} \label{eq:rdj}
j=\arg \max  _{j\in N_{i} } \, \left\{ \sum\limits_{u\in N_{i}} D(i,j,u)\quad \forall j\neq u\right\} 
\end{equation}

Else

\begin{equation} \label{eq:rdp}
P(i,j)=\left\{ \begin{array}{ll} 
\frac {\displaystyle \left(\sum\limits_{u\in N_{i}} D(i,j,u)\right)+1} { \displaystyle \sum\limits_{k\in N_{i}} \left(\left(\sum\limits_{u\in N_{i}}D(i,k,u)\right)+1\right) } & \quad if \, j\in N_{i} \land j\neq u \land k\neq u\\
\\
\\
0 & \quad otherwise\\
\end{array} \right.
\end{equation}

where all the parameters are the same as in Equations \ref{eq:rtj} and
\ref{eq:rtp}. This rule chooses the next node \textit{j} in the path
(when an ant is placed at node \textit{i}) considering the number of
neighbours dominated  for every one. So if \textit{q} $\le$
\textit{q$_{0}$}, the node which dominates more of the other
neighbours is chosen, otherwise the probability  
roulette wheel is used. In Equation \ref{eq:rdp} we add 1 to avoid a 0 
probability if no one dominates other neighbour.

There are two \textit{Evaluation Functions} (one per objective, 
again):

\begin{equation} \label{eq:funf}
F_{f} (Psol)=\sum\limits_{n\in Psol} [Cr(n-1,n)+\omega _{Ffo} \cdot
(1-ZO(n))]
\end{equation}

\begin{equation} \label{eq:funs}
F_{s} (Psol)=\sum\limits_{n\in Psol} [Ce(n-1,n)+\omega _{Fso} \cdot
(1-ZO(n))]
\end{equation}

Where \textit{Psol} is the solution path to evaluate and \textit{$\omega _{Ffo}$} and \textit{$\omega _{Fso}$} are weights related to the importance of visibility of the cells in the path. In Equation \ref{eq:funf} its importance will be small, it is less important to hide in a fast path and it will be high in Equation \ref{eq:funs} for the opposite reason. The other terms are the same that in Equations \ref{eq:nf} and \ref{eq:ns}.

Since CHAC is an ACS, there are two levels of pheromone updating, 
local and global, which update two matrices at each level. The 
equations for \textit{Local Pheromone Updating} (performed 
when a new node \textit{j} is added to the path an ant is building) are:

\begin{equation} \label{eq:lfuf}
\tau _{f} (i,j)=(1-\rho )\cdot \tau _{f} (i,j)+\rho \cdot \tau _{0f}
\end{equation}

\begin{equation} \label{eq:lfus}
\tau _{s} (i,j)=(1-\rho )\cdot \tau _{s} (i,j)+\rho \cdot \tau _{0s}
\end{equation}

Where \textit{$\rho$} in [0,1] is the common evaporation factor 
and \textit{$\tau _{0f}$}, \textit{$\tau _{0s}$} are the initial amount of pheromone in every edge for every objective, respectively:

\begin{equation} \label{eq:t0f}
\tau _{0f} = \frac {1} {(numc \cdot MAX_R)} 
\end{equation}

\begin{equation} \label{eq:t0s}
\tau _{0s} = \frac {1} {(numc \cdot MAX_E)}
\end{equation}

With \textit{numc} as the number of cells in the map to solve, \textit{MAX$_{R}$} is the maximum amount of resources going through a cell may require, and \textit{MAX$_{E}$} is the maximum energy cost going through a cell may produce (in the worst case).

The equations for \textit{Global Pheromone Updating} are:

\begin{equation} \label{eq:gfuf}
\tau _{f} (i,j)=(1-\rho )\cdot \tau _{f} (i,j)+\rho /F_{f}
\end{equation}

\begin{equation} \label{eq:gfus}
\tau _{s} (i,j)=(1-\rho )\cdot \tau _{s} (i,j)+\rho /F_{s}
\end{equation}

Only the solutions inside the Pareto set will make the global pheromone 
updating once all the ants have finished of building paths in every iteration.

The \textit{pseudocode} for CHAC is as follows:

\begin{verbatim}
Initialization of pheromone matrices with T0f and T0s

For i=1 to NUM_iterations

   For a=1 to NUM_ants

      ps=build_path(a)   /* Equations 1,2, [(3,4) or (6,7,8,9,10)] , 13,14 */

      evaluate(ps)   /* Equations 11,12 */

      if ps is non-dominated
         insert ps in Pareto_Set
         remove from Pareto_Set dominated solutions
      endif

   EndFor

   global_pheromone_updating   /* Equation 17,18 */

EndFor
\end{verbatim}

\section{Experiments and Results}
\label{sec:exp}

We would like to first emphasize that the values for the parameters and weights 
affect the results because parameters guide the search and balance 
the exploration and exploitation of the algorithm, and weights set 
the importance of every term in the heuristics and evaluation 
function. We fine-tuned both in order to obtain a good behaviour 
in almost all the maps; the values found were \textit{$\alpha$} = 1, \textit{$\beta$} = 2, \textit{$\rho$} = 0.1 and \textit{q$_{0}$} = 0.4, in the three experiments we show in this section. The last value establishes a balance between exploration and exploitation, tending to exploration (but without
abandoning exploitation altogether). The weights described in Equations 
1, 2, 11 and 12 are set to give more importance to minimizing 
distance to target point and consumption of resources in the 
speed objective, and to give more importance to minimizing visibility 
and consumption of energy in the safety objective.

The user can only decide the value for \textit{$\lambda$} parameter, 
which gives relative importance to one objective over the other, 
so if it is near 1, finding fastest path would be more important 
and if it is near 0, the other way round.

As every MO algorithm, CHAC yields a set of non-dominated solutions from 
which the user chooses using his own criteria, since, usually he only wants 
one solution. But in this algorithm the resulting Pareto set
is small (about five to ten different solutions on average, depending 
on the map size) because it only searches in the region of the 
ideal Pareto front delimited by the \textit{$\lambda$} parameter. We 
are going to represent in the results only `the best' solution inside all 
the Pareto sets of 30 executions, considering better the solution 
with the smallest cost in the most significant criteria (depending 
on \textit{$\lambda$} value). In order to clarify this concept, there 
is not a best path, but paths enough good from the military point 
of view. These paths have been selected by the military participation 
of the project taking into account tactical considerations and 
the features of every battlefield. 

CHAC have been implemented 
using the two state transition rules as two different algorithms.
We are going to compare the results between both CHAC implementations 
(with the same parameter values), and both of them with a greedy 
approach which uses the same heuristic functions as cost functions 
(the {\em pheromones} has no equivalent in this algorithm), but using
a \textit{dominance criterion} to select the next node of the actual
cell neighbourhood (it selects as next the node which dominates most
of the others  
neighbours considering their cost in both objectives). This algorithm
is quite simple, and sometimes does not even reach a solution
because it gets into a loop situation (back and forth between 
a set of cells consuming all the resources or energy).

We have performed experiments on three different maps; two of them
have cells of a single type of terrain (with different heights) in
order to avoid problems of visualization due to the black and white
figures. We executed every CHAC implementation 30 times with each
extreme value for \textit{$\lambda$} parameter (0.9 and 0.1) in order
to search for the fastest and the safest path (they cannot be 1
and 0 because always the  
two objectives must be considered). In every execution we chose 
the best solution (enough good from a military point of view) 
considering the objective we are minimizing and we make mean 
and standard deviation of them.

\subsection{River and Forest Map}

\begin{figure}[h]
\begin{center}
\includegraphics[bb = 0 0 930 450, scale=0.4]{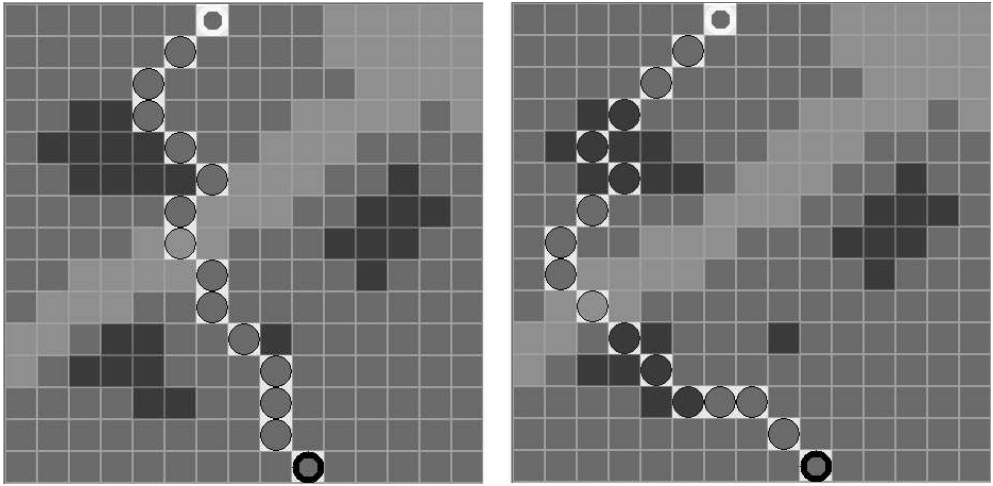}
\caption{ River and Forest map (forest cells in dark gray and water cells in light gray). Fastest (left) and safest 
(right) paths found. Both use forest cells to avoid being seen, but 
the latter goes through them to optimize safety.\label{fig:rfm}}
\end{center}
\end{figure}

The first scenario is a 15x15 map with a 
river flowing between the start and the target point and a patch of
forest. There is no enemy. The best of fast paths found and the best
of safe paths found are shown in Figure \ref{fig:rfm}, marked by 
circles.

\begin{table}[h]
\centering
\begin{tabular}{|l|c|c|c|c|c|c|c|c|c|}
\hline
& \multicolumn{4}{|c|}{\textit{Combined State Transition Rule}} & \multicolumn{4}{|c|}{\textit{Dominance State Transition Rule}} & 
\multirow{3}*{\em Greedy}\\
\cline{2-9}
& \multicolumn{2}{|c|}{Fastest (\textit{$\lambda$} = 0.9)} 
& \multicolumn{2}{|c|}{Safest (\textit{$\lambda$} = 0.1)}
& \multicolumn{2}{|c|}{Fastest (\textit{$\lambda$} = 0.9)} 
& \multicolumn{2}{|c|}{Safest (\textit{$\lambda$} = 0.1)} &  \\
\cline{2-9}
& $F_f$ & $F_s$ &$F_f$ & $F_s$ &$F_f$ & $F_s$ &$F_f$ & $F_s$ & \\
\hline
\; Best & \textbf{22.198} &  85.768 &  25.704 &  \textbf{61.189} &
\textbf{22.214} & 80.075 & 27.274 & \textbf{61.833} & \multirow{3}*{NO
SOLUTION}\\
\cline{2-9}
\; Mean & \textbf{22.207} & 85.941 & 28.505 & \textbf{67.296} & \textbf{22.595} & 88.405 & 28.724 & \textbf{66.365}& \\
& \textbf{$\pm$0.022} & $\pm$0.436 & $\pm$1.993 & \textbf{$\pm$3.360} & \textbf{$\pm$0.555} & $\pm$3.127 & $\pm$2.637 & \textbf{$\pm$2.881} &\\
\hline
\end{tabular}
\caption{Results for River and Forest map. (500 iterations, 20 ants) \label{tab:trf} }
\end{table}

As it can be seen in Table \ref{tab:trf}, the cost of the solutions for the 
fastest path have a small standard deviation which means possibly 
CHAC has reach a quasi-optimal solution (small search space and enough 
iterations and ants to solve it). Mean and standard deviation 
in the objective which is not being minimized are logically worse, 
because it has little importance. DSTR implementation yields similar 
results even improving sometimes mean and standard deviation values 
in main objectives so, its solutions are more similar between 
executions (the algorithm has robustness).

Figure \ref{fig:rfm} (left) shows the fastest path found, which goes zigzagging, 
because the algorithm usually moves to the most hidden cells when 
there are no enemies (in a radius that can be set up by the user 
and which is 10 in this experiment), even in the search for the 
fastest path, but with less relevance. Forest cells obstruct the 
line of sight so the unit tends to move near them while goes rather 
directly to the target point. Figure \ref{fig:rfm} (right) shows the safest 
path found which moves in a less straightforward way and hides by 
moving near, and even inside forest cells (hidden is very important 
in safe paths) although it mean bigger resource cost (forest 
cells have more resource cost assigned than flat terrain).

\subsection{Flat Terrain with Walls Map}

\begin{figure}[h]
\begin{center}
\includegraphics[bb = 0 0 930 450, scale=0.4]{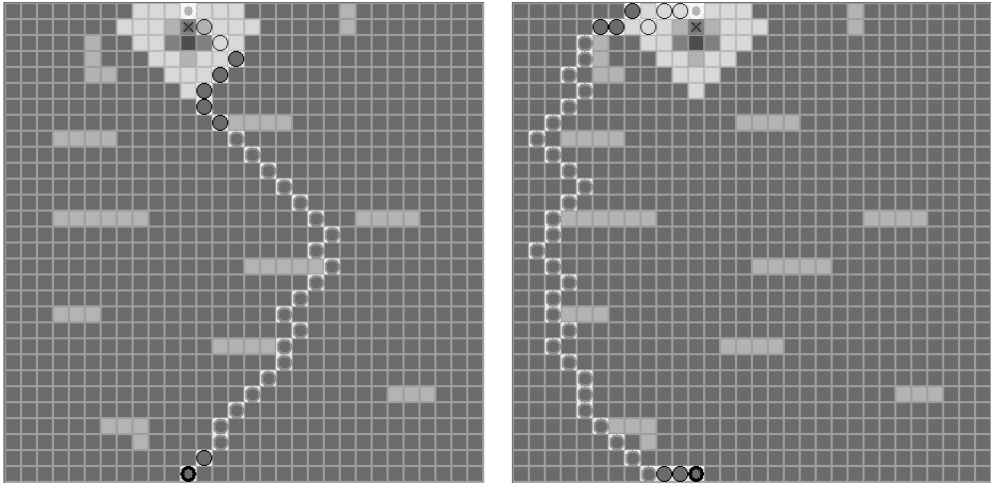}
\caption{Flat Terrain with Walls, and one enemy shooting 
in a zone near the target point (the enemy is marked with 'X', the
cells surrounding it in light shades of gray are the zone of weapons
impact and the other cells in light gray are the walls). Fastest
(left) and safest (right) paths found by CHAC. \label{fig:ftwm}}
\end{center}
\end{figure}

The second map contains 30x30 cells, and represents a flat terrain
with some `walls', there is one enemy unit protecting the target cell
(there are some cells affected by its weapons fire). The best of fast
paths found and the best of safe paths found are shown in Figure
\ref{fig:ftwm}. Cells with light gray border in the path are hidden to
enemy.

\begin{table}[h]
\centering
\begin{tabular}{|l|c|c|c|c|c|c|c|c|c|c|}
\hline
& \multicolumn{4}{|c|}{\textit{Combined State Transition Rule}} & \multicolumn{4}{|c|}{\textit{Dominance State Transition Rule}} & 
\multicolumn{2}{|c|}{\multirow{2}*{\textit{Greedy}}} \\
\cline{2-9}
& \multicolumn{2}{|c|}{Fastest (\textit{$\lambda$} = 0.9)} 
& \multicolumn{2}{|c|}{Safest (\textit{$\lambda$} = 0.1)}
& \multicolumn{2}{|c|}{Fastest (\textit{$\lambda$} = 0.9)} 
& \multicolumn{2}{|c|}{Safest (\textit{$\lambda$} = 0.1)} 
& \multicolumn{2}{|c|}{} \\
\cline{2-11}
& $F_f$ & $F_s$ &$F_f$ & $F_s$ &$F_f$ & $F_s$ &$F_f$ & $F_s$ &$F_f$ & $F_s$ \\
\hline
\; Best & \textbf{36.500} &  133.000 &  38.000 &  \textbf{112.700} &
\textbf{39.000} & 133.100 & 48.500 & \textbf{113.400} &
\multirow{3}*{46.7} & \multirow{3}*{322.9} \\
\cline{2-9}
\; Mean & \textbf{37.220} & 141.973 & 45.650 & \textbf{121.210} & \textbf{43.350} & 240.660 & 59.530 & \textbf{123.980} & & \\
& \textbf{$\pm$1.216} & $\pm$2.904 & $\pm$3.681 & \textbf{$\pm$10.425} & \textbf{$\pm$1.592} & $\pm$85.254 & $\pm$10.174 & \textbf{$\pm$8.119} & & \\
\hline
\end{tabular}
\caption{Results for the Flat Terrain with Walls map. (1000 iterations, 30 ants) \label{tab:tftw} }
\end{table}

As it can be seen in Table \ref{tab:tftw}, CHAC (in the two
implementations) outperforms the greedy algorithm by almost 25\% in
resource cost and 75\% in safety costs. The reason for this is that
the greedy path is rather straight, which means  
low resource expenses; but, at the same time, it means the enemy 
sees the unit all the time (it moves through uncovered cells) 
and the safety cost is dramatically increased (it depends too 
much on cell visibility). The cost for fast path, which also depends
on visibility, is increased too. The standard deviation is 
small in CSTR, but it grows a little in values of safest paths
which could mean a bigger exploitation factor is needed. Again, 
in the DSTR implementation, results are similar to those obtained with
CSTR, but with better standard deviation; however, its performance is poorer
(even quite bad in one case) for the objectives it is not minimizing, 
which supports the theory that a higher exploitation level is needed 
in this scenario.

In Figure \ref{fig:ftwm} (left), we can see the fastest path found; the unit moves 
in a curve mode in order to hide from enemy behind the walls 
(it consider the hidden of cells from the enemy cell). It surrounds 
this terrain elevations because resource cost of going through 
them is big. On the other hand, in Figure \ref{fig:ftwm} (right) we can see 
the safest path found which moves surrounding the left side walls 
but remaining during more cells hide for the enemy. In both cases, 
the unit moves inside the zone affected by weapons choosing the 
less damaging cells, even in the safe case it arrives by a flank 
of the enemy unit where they have less fire power and this inflict 
less damage. It represents a good behaviour, very tactical in 
the case of safest path because attack to an enemy by the flank is 
one principle of military tactics.

\subsection{Valleys and Mountains Map}

\begin{figure}[h]
\begin{center}
\includegraphics[bb = 0 0 930 450, scale=0.4]{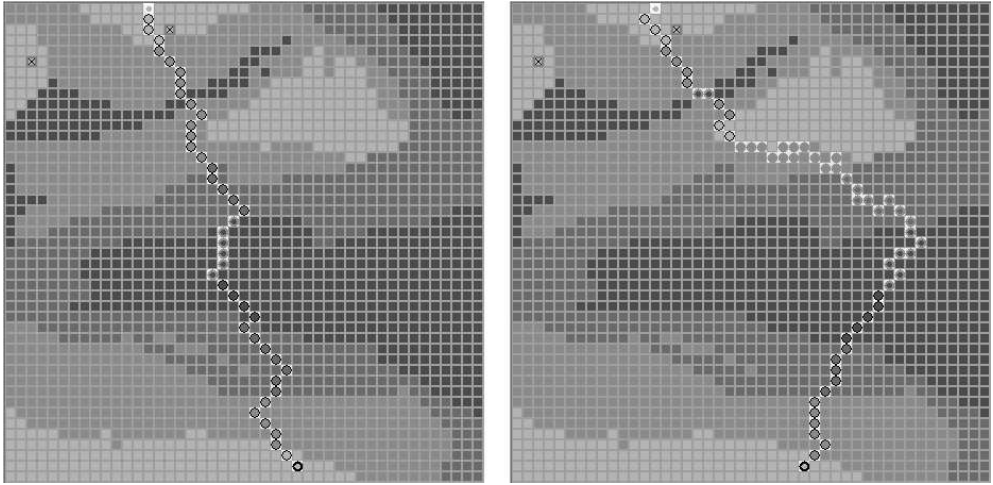}
\caption{Valleys and Mountains, with 2 enemy units (marked with 'X') on 
watch. Fastest (left) and safest (right) path found. \label{fig:vmm}}
\end{center}
\end{figure}

This 45x45 map represents some big terrain accidents, with clear zones
representing peaks (the more higher the more clearer) and dark zones
representing valleys or cliffs. There are two enemies located in both
peaks and the target point is behind one of them. The best of fast
paths found and the best of safe paths found are shown in Figure
\ref{fig:vmm}. Circle cells mark the paths, those with light gray border in the path are hidden to both enemies.

\begin{table}[h]
\centering
\begin{tabular}{|l|c|c|c|c|c|c|c|c|c|}
\hline
& \multicolumn{4}{|c|}{\textit{Combined State Transition Rule}} & \multicolumn{4}{|c|}{\textit{Dominance State Transition Rule}} & 
\multirow{3}*{\textit{Greedy}} \\
\cline{2-9}
& \multicolumn{2}{|c|}{Fastest (\textit{$\lambda$} = 0.9)} 
& \multicolumn{2}{|c|}{Safest (\textit{$\lambda$} = 0.1)}
& \multicolumn{2}{|c|}{Fastest (\textit{$\lambda$} = 0.9)} 
& \multicolumn{2}{|c|}{Safest (\textit{$\lambda$} = 0.1)} & \\
\cline{2-9}
& $F_f$ & $F_s$ &$F_f$ & $F_s$ &$F_f$ & $F_s$ &$F_f$ & $F_s$ &  \\
\hline
\; Best & \textbf{70.500} &  334.600 &  84.500 &  \textbf{285.800} &
\textbf{73.500} & 374.300 & 76.000 & \textbf{354.600} &
\multirow{3}*{NO SOLUTION}\\
\cline{2-9}
\; Mean & \textbf{75.133} & 357.800 & 105.517 & \textbf{311.390} &
\textbf{77.950} & 397.280 & 88.780 & \textbf{371.890} &\\

& \textbf{$\pm$2.206} & $\pm$9.726 & $\pm$17.138 & \textbf{$\pm$19.541} & \textbf{$\pm$1.724} & $\pm$11.591 & $\pm$13.865 & \textbf{$\pm$7.522} &\\
\hline
\end{tabular}
\caption{Results for Valleys and Mountains map. (2000 iterations, 70 ants) 
\label{tab:tvm} }
\end{table}

As Table \ref{tab:tvm} shows, both implementations yield
a low standard deviation, which means the algorithm is robust with
respecto to initialization. CSTR implementation  
results may be nearer an optimal path, but in the DSTR case, the
solutions have big costs (compared with the others), which means it
needs more iterations or a greater exploitation factor to improve them. As 
in previous experiments, mean and standard deviation in the objectives 
that the algorithm is not minimizing are worse, because they 
have little importance. In this case, the differences between 
the cost of an objective when search minimizing this objective 
and when search minimizing the other are bigger than in the previous 
maps because the search space is bigger too.

In Figure \ref{fig:vmm} (left) we can see the fastest path found, the unit 
goes in a rather straight way to the target point, but without 
considering the hidden of cells (from the position of both enemies) so 
the safety cost increases dramatically. On the other hand, in 
Figure \ref{fig:vmm} (right) we can see the safest path found which represents 
a curve (distance to target point has little importance) which 
increases speed cost, but the unit goes through many hidden cells. 
This behaviour is excellent from military tactical point of view.

\section{Conclusions and Future Work}
\label{sec:conc}

In this paper we have described a MOACO algorithm called CHAC which
tries to find the fastest and safest path, whose relative importance is set by the user, for a simulated military
unit. This algorithm can use different state transition rules, and, in
this papeer, two of them have been presented and tested. The first 
one combines heuristic and pheromone information of two objectives 
(Combined State Transition Rule, CSTR) and the second one is based 
on dominance over neighbours (Dominance State Transition Rule, 
DSTR).

The algorithm using both state transition rules has been tested in
several different scenarios yielding very good results in a subjective
assessment by the military  
staff of the project (Mr. Millán and Torrecillas) and being perfectly compatible with military 
tactics. They even offer good solutions in complicated maps 
in less time than a human expert would need. Moreover it is possible to 
observe an inherent emergent behaviour studying the solutions 
because it tends to be similar to those  a real 
commander would, in many cases, take. In addition CHAC (using any
state transition rule of the ones tested so far) outperforms 
a greedy algorithm. In the comparison between them, CHAC with CSTR 
yields better results in the same conditions, but CHAC with DSTR is
more robust, yielding solutions that perform similarly, independently
of the random initial conditions. If 
we increase exploitation level or iterations of algorithm, DSTR
approach offers similar results to CSTR. 

As future work, we will compare CHAC with path finding 
algorithms better than the simple greedy we have used here. From the algorithmic point 
of view, we will try to approach more systematically parameter 
setting, and investigate its performance in dynamic environments 
where, for instance, the enemy can move and shoot on sight. We 
will also try to evaluate its performance in environments with 
a hard time constraint, that is, in environments where the algorithm
cannot run for an unlimited amount of time, but a limited and small
one, which is usually the case in real combat situations. From the
implementation point of view,  
it would be interesting to implement it within a real combat 
training simulator, that includes realistic values for most variables 
in this algorithm, including fuel consumption and casualties 
caused by projectile impact.

We will also try to approach scenario design more systematically,
trying to describe its difficulty by looking at different
aspects. This will allow us to assess different aspects of the
algorithm and relate them to scenario difficulty, finding out which
parameter combination is better for each scenario.

\section*{Acknowledgements}
This work has been developed 
within the SIMAUTAVA Project, which is supported by Universidad 
de Granada and MADOC-JCISAT of Ejército de Tierra de
España. Mr. Millán is a Lieutenant Colonel and Mr. Torrecillas is a
Major of the Spanish Army Infantry Corps.

\bibliographystyle{unsrt}
\bibliography{CHAC}

\begin{thebibliography}{1}

\bibitem{dori99a}
M.~Dorigo and G.~Di Caro.
\newblock The ant colony optimization meta-heuristic.
\newblock In D.~Corne, M.~Dorigo, and F.~Glover, editors, {\em New Ideas in
  Optimization}, pages 11--32. McGraw-Hill, 1999.

\bibitem{dori2003}
M.~Dorigo and T.~Stützle.
\newblock The ant colony optimization metaheuristic: Algorithms, applications,
  and advances.
\newblock In G.A.~Kochenberger F.~Glover, editor, {\em Handbook of
  Metaheuristics}, pages 251--285. Kluwer, 2002.

\bibitem{Coello2002}
Carlos A.~Coello Coello, David A.~Van Veldhuizen, and Gary~B. Lamont.
\newblock {\em Evolutionary Algorithms for Solving Multi-Objective Problems}.
\newblock Kluwer Academic Publishers, 2002.

\bibitem{bicriteria}
C.~García-Martínez, O.~Cordón, and F.Herrera.
\newblock An empirical analysis of multiple objective ant colony optimization
  algorithms for the bi-criteria {TSP}.
\newblock In {\em ANTS 2004. Fourth International Workshop on. Ant Colony
  Optimization and Swarm Intelligence}, number 3172 in LNCS, pages 61--72.
  Springer, 2004.

\bibitem{IMM2001}
Steffen Iredi, Daniel Merkle, and Martin Middendorf.
\newblock Bi-criterion optimization with multi colony ant algorithms.
\newblock In E.~Zitzler, K.~Deb, L.~Thiele, C.~A.~Coello Coello, and D.~Corne,
  editors, {\em Proceedings of the First International Conference on
  Evolutionary Multi-Criterion Optimization {(EMO 2001)}}, volume 1993 of {\em
  Lecture Notes in Computer Science}, pages 359--372, Berlin, 2001.
  Springer-Verlag.

\bibitem{Baran}
B.~Barán and M.~Schaerer.
\newblock A multiobjective ant colony system for vehicle routing problem with
  time windows.
\newblock In {\em IASTED International Multi-Conference on Applied
  Informatics}, number~21 in IASTED IMCAI, pages 97--102, 2003.
\newblock
  \url{http://www.scopus.com/scopus/inward/record.url?eid=2-s2.0-1442302509&pa%
rtner=40&rel=R4.5.0}.

\end{thebibliography}

\end{document}